# Fog Computing as Privacy Enabler

Frank Pallas*, Philip Raschke*, David Bermbach*° – *TU Berlin, °ECDF

Despite broad discussions on privacy challenges arising from fog computing, the authors argue that privacy and security requirements might actually drive the adoption of fog computing. They present four patterns of fog computing fostering data privacy and the security of business secrets, complementing existing cryptographic approaches. Their practical application is illuminated on the basis of three case studies.

## Introduction

Fog computing is an emerging computing paradigm in which applications and data are no longer deployed in the cloud alone. Instead, the cloud is enhanced with machines running closer to end users, e.g., in mobile applications, or data producers, e.g., in Internet of Things (IoT) scenarios [1].

The typical drivers of this move towards fog computing are latency and bandwidth: Application domains such as autonomous driving or 5G mobile applications call for low latency responses. Here, a local edge server is able to respond much faster than a remote cloud server. The IoT, in turn, is inherently limited by bandwidth constraints as the data output from large scale sensor deployments can quickly exceed the available bandwidth to the cloud. In that case, preprocessing at the edge can significantly reduce the amount of data that needs to be sent over the wire.
While these two drivers are intuitive and, thus, usually the first ones mentioned when motivating fog computing, there is another driver: privacy. In light of recent legislation, i.e., the General Data Protection Regulation (GDPR), privacy will, in our opinion, be one of the most dominant drivers of fog computing. Nevertheless, positive effects of fog computing for privacy have largely been neglected so far. To the contrary, this paper is actually inspired by an earlier article in this magazine which claimed that privacy might hamper broad fog adoption [2]. While we agree that there are indeed some privacy (and security) risks in fog computing, we believe that these are largely technical challenges that can be surmounted with existing technology. On the other hand, fog computing provides invaluable privacy opportunities that are primarily of structural nature and can valuably complement existing, especially cryptographic techniques such as homomorphic encryption [7], secure multi-party computation [9], or mechanisms for differential privacy [10]. These structure-based opportunities can be used to significantly heighten data privacy, as compared to well-established cloud-centric models of data processing. In this paper, we give an overview of these privacy opportunities and demonstrate their potential through a number of case studies.

## WHAT IS FOG

Fog computing is the combination of cloud and edge computing enhanced with compute capacity in the network between edge and cloud. On an abstract level, this means that the centralized, homogeneous cloud is replaced with a geo-distributed, highly heterogeneous compute environment where compute infrastructure may be controlled by multiple entities. This implies that the fog can be seen as the natural evolution of hybrid clouds: public and private clouds are still used, but are augmented with large numbers of additional infrastructure elements in the network between public and private cloud as well as at locations even closer to the edge. These elements come in a variety of flavors ranging from single board computers to small clusters and are operated by a multitude of providers [1].

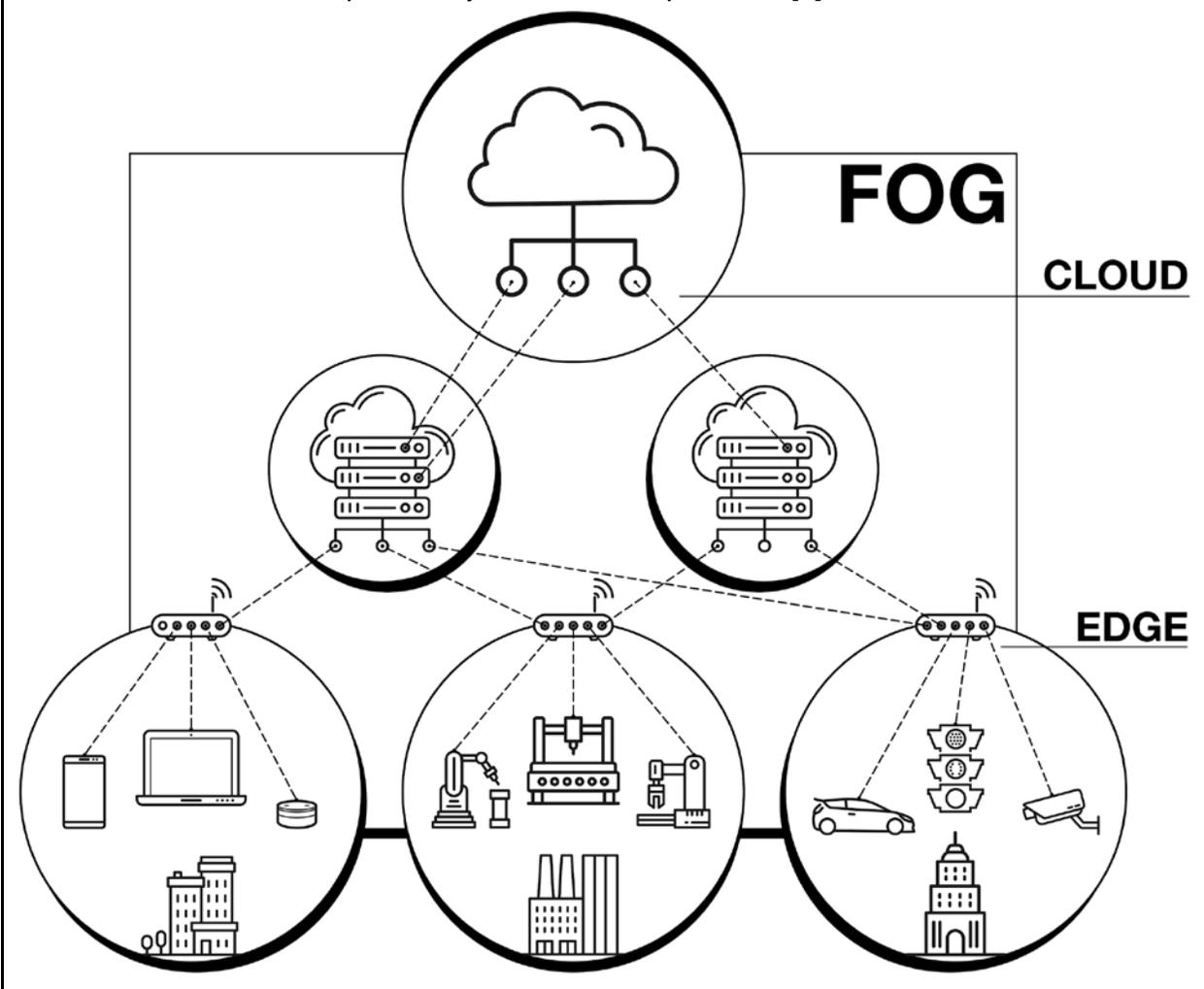

Figure 1: Fog computing combines and integrates cloud and edge computing with additional compute capacity in the network in-between, leading to highly distributed and heterogeneous environments.

> PRIVACY and BUSINESS SECRETS
>
> Challenges with regard to privacy and business secrets often overlap. In particular, this is the case for questions about who is able to access certain data in cloud computing settings. The computational overheads raised by approaches like fully homomorphic encryption – which are often proposed for such use cases [5, 7] – are still prohibitively high for most realistic scenarios [4]. Even though data can be encrypted during transit and while at rest, cloud providers thus typically need access to unencrypted data as soon as reasonable functionality is desired. Unable to actually exert control over the access to and usage of such data, a company employing cloud services for handling customer data therefore has to blindly trust the legal compliance of the cloud or SaaS provider. At the same time, it has to bear the ultimate risk of being held accountable in case of misbehavior resulting in high fines.
>
> A comparable challenge arises with regard to business secrets in case of using cloud-based services for processing sensitive internal data, such as production volume, material flow, or, in particular, large-volume monitoring and control data in the context of industry 4.0. Again, any company employing cloud-based services for such purposes has to trust in the proper handling on the side of the provider.
>
> Aside from such aspects of data secrecy, privacy regulations like the GDPR also raise additional requirements, ranging from data minimization obligations and purpose limitation to the provision of transparency and geographical limitation of storage and processing which go far beyond what cryptographic techniques can offer . Following the paradigm of privacy/data protection by design, all these and many more requirements need to be addressed using technical measures.

# Opportunities

In particular, we see four structural patterns of applied fog computing that provide significant benefits in matters of data privacy but also for the handling of business secrets. These patterns are not necessarily new and may already be known from, e.g., hybrid clouds. In fog computing, however, they become inherent and fundamental principles of application architectures.

**1: Service Execution On-Premises**
A core functionality of fog computing is the movement (or: relocation) of data storage and processing away from centralized entities like cloud datacenters to on-premises edge servers. Basically, fog computing allows to decouple the software part of a SaaS business from the execution environment this software runs on. While a software executable is still provided as a service, it is executed in an environment (including hardware as well as the underlying software stack) which remains under the control of the service customer. In contrast to cloud-based models, customers therefore no longer need to completely hand-over control over their data to the service provider for being able to use said service. This, in turn, provides significant benefits for privacy as well as for the protection of business secrets. This pattern can already be found in practice. Amazon, for instance, offers ready-to-use server racks which are installed as on-premises edge servers under the physical control of the customer. These "Outpost" installations, however, run standard Amazon cloud services managed by the provider.

**2: Multi-Staged Filtering**

Data minimization is one of the core principles of privacy. This corresponds nicely to fog scenarios where data created near the edge is no longer streamed to the cloud for, e.g., bandwidth reasons. Instead, it is processed across multiple stages, starting at the edge, with only aggregated values arriving in the cloud. In a typical IoT scenario, for instance, sensor data is processed at the edge to either aggregate data or to filter out and discard irrelevant data items. Depending on the use-case, its data requirements, and the computational constraints it raises, this filtering might also comprise mechanisms providing local differential privacy [5, 6]. The same happens at later filtering stages within the fog before eventually arriving in the cloud for final processing. Every stage reduces the amount of data permanently stored in the cloud which will be accessible for the respective cloud providers. The pattern therefore allows to minimize the amount of data accessible for the different parties involved to the necessary minimum. In the context of business secrets, this reduces the risk of disclosure while in the context of privacy, it allows for compliance with the principle of data minimization, which is mandatory in most respective regulations such as the GDPR.

**3: Decoupled Data Hubs**

One of the main privacy risks in cloud computing is the correlation of data which was collected for independent purposes. Nowadays, such correlation of data – whether accidental or malicious – has become simple due to the wide availability of big data technology and the increasing accumulation of data in consolidated cloud data stores. Even though these cloud data stores typically employ sophisticated access management systems, these do not solve the underlying structural risk emanating from large centralized data lakes. When data, however, is not stored in a centralized cloud but rather distributed geographically across a multitude of fog nodes operated by different providers, joins of data sets (even accidental) become incredibly expensive. Fog-based decoupled data hubs cannot avert data correlation and inference completely (data correlation for single entities may still be possible), but it is no longer feasible to do that on a large scale. Using federated learning approaches [11], decoupled data hubs, however, still support advanced machine learning use cases such as predictive maintenance across multiple machine users. Overall, decoupled data hubs clearly counteract privacy intrusion as well as provider-side fiddling with internal business secrets.

**4: Fine-Granular Control of Data Placement**

Privacy legislation mandates that data may not move to certain countries. Business secrets, in turn, shall not leave company premises, and users may feel more comfortable when they can actually know and control the place where their data is stored. This is in stark contrast to the cloud which through its virtualized nature here provides only limited transparency and control. In fog computing, in contrast, the latency and bandwidth implications of geo-distribution call for explicitly exposing data placement to applications. Furthermore, the sheer number of edge locations is an enabler for really fine-grained geo-placement [12]. Respective strategies (see, e.g., [8]) can also be used to satisfy - going beyond performance- and QoS-related requirements - privacy-and security- related obligations and constraints. Furthermore, users may also welcome the option to place their data with only a subset of fog providers. All these aspects together provide the means to let application developers better meet regulatory demands and to allow users to manage the physical location of their data.

# Case Studies

To illustrate how the four identified fog usage patterns help to improve privacy or handling of sensitive business secrets, consider the following example use cases:

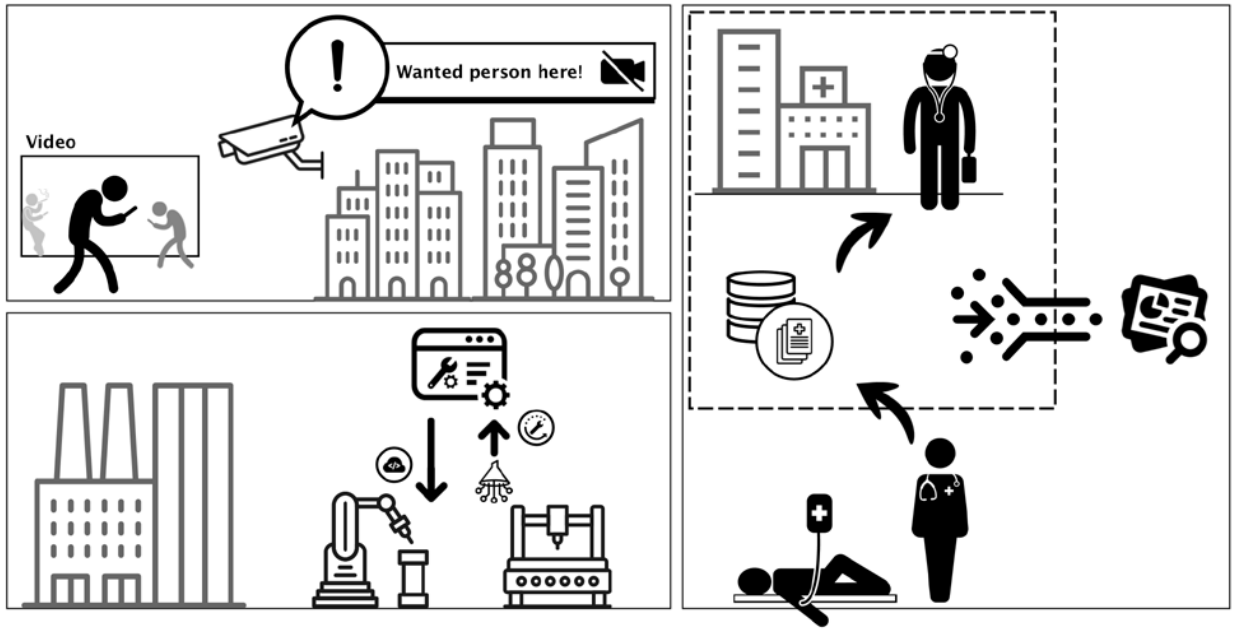

Figure 3: Fog Computing may provide data privacy and security benefits in such diverse scenarios as video-surveillance, predictive maintenance, or eHealth.

**Privacy-Friendly, Fog-Based Video Surveillance**
Video surveillance is a widely used approach for improving safety in public places. Typically, the resulting video feed is sent to the cloud where it is processed and stored indefinitely. In [3], we presented an alternative approach where the video stream is analyzed at the edge. Using facial recognition models pre-trained in the cloud, only the video snippets containing persons of interest are sent to the cloud. All other video data is buffered at the edge for a few days and may only be retrieved through a well-controlled process. Both, the face search requests as well as extraordinary data retrieval require a court order (ensured, e.g., through a four eyes procedure) and a blockchain-inspired audit trail is created to avoid misuse. As the necessary edge infrastructure, we used standard off-the-shelf single-board computers (Raspberry Pi) which were already powerful enough to process high-definition video streams unless these contained large groups of people. This shows that such an approach is indeed feasible for deployments in practice which would significantly improve privacy of citizens over the state-of-practice (cloud-based analytics).

In this case, the patterns of multi-staged filtering (2) and of decoupled data hubs (3) are employed. Here, filtering refers to the edge-side analysis of raw video footage and the reduction of transmitted data to those snippets actually containing persons of interest. Intermediate nodes could also have been used to filter data based on further criteria (e.g., reporting only certain matches to federal agencies). The temporary storage of full raw data at the edge for the carefully restricted case that further examination is necessary, in turn, resembles the pattern of decoupled data hubs – even though only temporary.

In established, cloud-based analytics, in contrast, the entire raw footage would be sent to a centralized cloud data store, structurally making it available for arbitrary analytics, especially across all covered locations. In such deployments, it is hence possible to create movement profiles of arbitrary citizens. Opposed to this, the above-sketched fog-based approach provides structural benefits from the perspective of data minimization as well as with regard to the risk of inappropriate or unintended correlation.

**Business Secrets in Vendor-Provided Predictive Maintenance**

One of the core applications in Industry 4.0 scenarios is predictive maintenance – the prediction of necessary maintenance based on fine-grained, high-volume operational data of industrial facilities and respective machine learning algorithms. For manufacturers of industrial production machines, predictive maintenance is commonly seen as a natural extension of their existing service portfolio as, in particular, they are best-suited to interpret observed operational data. On the other hand, owners of the machines may fear the exposure of their internal operational data and therefore hesitate to participate in this – otherwise advantageous – model.

With fog computing, in turn, manufacturer-provided services for maintenance prediction may be distributed across machine-attached devices and the company's private cloud, both controlled by the machine owner, without sending detailed information on internal production processes to external parties.

We implemented this model in collaboration with a manufacturer of digitally programmable high-precision machines. This allows for predictive maintenance based on local analytics of high-volume monitoring data being executed on the customers' premises without revealing detailed production data to the machine-manufacturer. Instead, only well-selected aggregates are sent to the manufacturer in specific cases. Beyond maintenance-related analysis results, this can (depending on the agreement between customer and machine-manufacturer) also include, e.g., precalculated analysis results covering the last 24 hours to improve the prediction models. Per machine, this required one node of the usual desktop class. An alternative approach, following the same principles, would have been to employ federated learning methods such as [11] which allow different entities to jointly train a (federated) model without exposing their individual training data possibly containing business secrets.

In this case, we employed the above-mentioned patterns of service execution on-premises (1), filtering (2), and decoupled data hubs (3). Compared to the alternative model of having all data sent to a centralized vendor-cloud, the customer is significantly less exposed to the risk of having their internal production data inappropriately analyzed or even exploited. Also, the necessary bandwidth could be significantly reduced without noteworthy limitation of functionality.

**eHealth and Data Sharing and Reuse**

The use of health-related data for different purposes by different actors is one of the main principles driving eHealth adoption. In particular, this provides benefits when multiple parties (e.g., multiple departments of a hospital or different hospitals or clinics of the same group, laboratories, etc.) are involved in the treatment of a single patient and can easily access existing examination data. However, regulations applicable to the health sector often raise strict constraints on where health-related data may be stored (within a given facility or a certain country).

As opposed to established models of cloud-based data sharing, compliance with such regulations can be significantly streamlined with fog computing's inherent capabilities for data placement control (pattern 4 above). Another often-mentioned use case for health related data regards the reuse of examination data (e.g., blood tests) for medical research. In this case, regulations strictly require respective data to be pseudonymized or anonymized, depending on the party conducting the research. For the automated application of such data minimization measures, data might – following the above-sketched pattern of multi-staged filtering – be preaggregated per patient at the edge, pseudonymized in a hospital data center, and finally anonymized through aggregation across patients in a hospital group data center before releasing it to external entities.

We are currently implementing this model together with a nation-wide hospital group in a European country through a novel data management and exchange platform. Early results are promising and already confirm the expected benefits in matters of easing the collaboration between different parties in line with existing regulations. Compared with an alternative model of a centralized cloud data store with traditional access restrictions, the sketched model allows us to clearly separate data of different levels of sensitivity and to clearly manage the storage locations of these data in line with regulations such as the GDPR or more specific national laws.

# Roundup

The case studies we presented highlight the privacy opportunities implied by fog computing. By relocating data storage and processing to the edge, applications can be executed on-premise, i.e., sensitive data of a natural person or business is never given away. If data transmission becomes indispensable, filters can be used to only disclose derived or aggregated data or to omit sensitive information. Thus, data privacy as well as business secrets are better protected. Furthermore, a distribution of data across multiple nodes makes it harder to aggregate it in order to reveal correlations. Consequently, the privacy-affecting effects of big data can be mitigated.

In all our case studies, users' data sovereignty is improved by the fog computing paradigm. Data owners, natural persons or businesses, are able to decide which information – personal data or business secrets – they want to disclose to whom, where it should be stored, and who has access to it. This provides significant benefits over established cloud-based environments, where users effectively give away all these choices at the moment they upload their data to the cloud.

Given these achievable advantages in matters of legal compliance as well as the alignment with business interests, fog computing might become a serious competitor for the traditional cloud computing paradigm. However, cumbersome set ups and configurations currently hamper its adoption [1]. The bar is raised high by cloud providers here, yet it is not out of reach. Fog computing can claim itself many advantages over the often problematic cloud computing paradigm. Its respect for data privacy is definitely one of them.

Frank Pallas is a senior researcher in the Information Systems Engineering research group of TU Berlin. His research interests particularly include interdisciplinary aspects of privacy and security in cloud-, fog-, and web-based computing as well as in the IoT. He holds a PhD with distinction in computer science from TU Berlin. Contact him at fp@ise.tu-berlin.de

Philip Raschke is a PhD student and researcher in the Service-centric Networking research group of TU Berlin. His research interests include data privacy and protection, Web and mobile tracking, and implementation of usable transparency-enhancing technologies. Contact him via philip.raschke@tu-berlin.de.

David Bermbach is an Assistant Professor and head of the Mobile Cloud Computing research group at TU Berlin and the Einstein Center Digital Future. His research interests include cloud service benchmarking but also novel compute and data management platforms for cloud and fog computing. He holds a PhD with distinction in computer science from Karlsruhe Institute of Technology. Contact him at db@mcc.tu-berlin.de.